\documentclass[11pt]{article}

\usepackage{graphicx}
\usepackage{float}
\usepackage{amssymb}

\usepackage[T1]{fontenc}
\usepackage[latin9]{inputenc}
\usepackage{array}
\usepackage{multirow}
\usepackage{amsmath}
\usepackage{setspace}
\usepackage{esint}
\usepackage[authoryear]{natbib}

\usepackage{amsthm}
\usepackage{thmtools}\usepackage{latexsym}

\newcommand{\openr}{\hbox{${\rm I\kern-.2em R}$}}
\newcommand{\openn}{\hbox{${\rm I\kern-.2em N}$}}

\usepackage{xcolor}

\usepackage{JASA_manu}
\usepackage{subcaption}

\usepackage{array}

\newcolumntype{+}{!{\vrule width 2pt}}

\newlength\savedwidth

\newcommand\thickhline{\noalign{\global\savedwidth\arrayrulewidth\global\arrayrulewidth 2pt}%
\hline
\noalign{\global\arrayrulewidth\savedwidth}}

\begin{document}

\title{Network Dependence Can Lead to Spurious Associations and Invalid Inference}
\author{Youjin Lee and Elizabeth L. Ogburn\thanks{Department of Biostatistics, Johns Hopkins Bloomberg School of Public Health, Baltimore, MD, USA}}
 \date{}
\maketitle



\begin{center}
\textbf{Abstract}
\end{center}

Researchers across the health and social sciences generally assume that observations are independent, even while relying on convenience samples that draw subjects from one or a small number of communities, schools, hospitals, etc. A paradigmatic example of this is the Framingham Heart Study (FHS). Many of the limitations of such samples are well-known, but the issue of statistical dependence due to social network ties has not previously been addressed.  We show that, along with anticonservative variance estimation, this can result in \emph{spurious associations due to network dependence}. Using a statistical test that we adapted from one developed for spatial autocorrelation, we test for network dependence in several of the thousands of influential papers that have been published using FHS data. Results suggest that some of the many decades of research on coronary heart disease, other health outcomes, and peer influence using FHS data may suffer from spurious associations, error-prone point estimates, and anticonservative inference due to unacknowledged network dependence. These issues are not unique to the FHS; as researchers in psychology, medicine, and beyond grapple with replication failures, this unacknowledged source of invalid statistical inference should be part of the conversation.


\vspace*{.3in}


\newpage

\section{Introduction \label{sec:Introduction}}

Whenever human subjects are sampled from one or a small number of communities, schools, hospitals, etc., as is routine in the health and social sciences, they may be connected by social ties, such as friendship or family membership, that could engender statistical dependence. This is \textit{network dependence}.  We show that when an outcome and an exposure of interest are causally and statistically independent from one another, but both exhibit network dependence, it can result in \emph{spurious associations due to dependence}.  Distinct from confounding, which usually arises when there is a common cause of an exposure and outcome, spurious associations due to dependence are unrelated to the causal relationships among the exposure, outcome, and other variables.  
Although this phenomenon is well-known in the econometrics literature on time series, it has never, to our knowledge, been addressed in a general dependent data framework.  We focus primarily on network dependence, but spurious associations due to dependence can arise no matter  the source of dependence across observations--be it spatial, genetic, network, temporal, or other.


In order to begin to assess how ignoring network dependence may impact research relying on convenience samples of interconnected subjects, we propose a test that can help detect when these might be a problem and apply it to data from the Framingham Heart Study (FHS), one of the few studies not explicitly about social networks for which some data on network ties is available.  The FHS is a paradigmatic example of an epidemiologic study comprised of individuals from a single tight-knit community, and it has served as a basis for a large literature on phenomena from heart disease to social contagion, all using statistical methods that assume independent and identically distributed (i.i.d.) data. Our results suggest that the i.i.d. assumption on which thousands of FHS papers have relied does not consistently hold, and therefore that spurious associations due to network dependence may be widespread among studies using FHS data, and likely beyond. 

Although the replication crisis in the health and social sciences has drawn attention to many ways that the flawed application of statistics can result in spurious findings, network dependence and spurious associations due to dependence have received little to no attention.
We focus throughout on p-values and null hypothesis testing, because they continue to dominate the applied literature and because they provide a precise framework for discussing issues like the validity of statistical inference, replication crises, and false positive rates. All of our examples and simulations involving two variables are constructed under the null of no association between them.  But most of our substantive points apply to point estimation and to other modes of inference as well.  We refer to a statistical inferential procedure as \emph{invalid} if it (a) rests on assumptions that are not true, (b) has a Type I error rate that is greater than the nominal rate, or (c) results in a confidence interval coverage rate that is less than the nominal rate.  

\section{Spurious associations due to dependence}

A predictor, $\mathbf{X}=X_1,...,X_n$, and an outcome, $\mathbf{Y}=Y_1,...,Y_n$, may appear to be associated when they are in fact independent from one another but share similar correlation structures across units.  That is, estimates of association between $X$ and $Y$ may be concentrated away from their null values, possibly even as $n$ goes to infinity.  We will refer to this phenomenon as \emph{spurious associations due to dependence}.  This phenomenon is well-known in the econometrics literature on time series, where it is referred to as spurious, nonsense, or volatile correlations \citep{phillips1986understanding,ernst2017yule}, and where much work has been done on pre-whitening data, e.g. extracting independent increments, before attempting to learn about associations between time series.  (See Section 12.2.3 of \citealp{kass2014analysis} for a simple illustration of spurious associations due to temporal autocorrelation in linear regression, and Section 12.4 of \citealp{efron2016computer} for a related discussion of \emph{ephemeral predictions}.) This phenomenon is also closely related to confounding due to cryptic relatedness and other population structure in genetic association studies  \citep{sillanpaa2011overview}, where SNPs and phenotypes of interest may exhibit common dependence structure, due, e.g., to familial structure among the subjects in the study.  Here too there has been considerable work done on modeling away the dependence within the variables of interest before assessing the association between them.  
	
Despite the fact that spurious associations can arise whenever $X$ and $Y$ are dependent across units, with similar dependence patterns for both $X$ and $Y$, we are not aware of any prior literature on this general phenomenon.  Indeed, this issue seems to be overlooked by both applied researchers and statisticians, outside of the contexts of time series and GWAS data. We will define and explain spurious correlations due to dependence before focusing on the specific case of network dependence.  We propose a test that can help researchers diagnose whether spurious correlations may be an issue in social network data, but as we discuss in Section \ref{sec:solutions}, there is not likely to be a one-size-fits all solution to the problem of spurious associations in networks.   

We use a toy example in order to convince the reader that the phenomenon of spurious associations due to dependence is real, and is entirely distinct from causal confounding. 
Begin with the premise that $\mathbf{X}=X_1,...,X_n$  and $\mathbf{Y}=Y_1,...,Y_n$ are causally and statistically independent from one another.  If, in addition, $X_1,...,X_n$ are independent and $Y_1,...,Y_n$ are independent, then valid tests of the absence of an association or correlation between $X$ and $Y$ are easy to come by.  But suppose instead that $X_1,...,X_n$ are dependent across observations, and so are $Y_1,...,Y_n$.  In particular, suppose that $\mathbf{X}$ is a series that is either increasing or decreasing in the unit ID $i$, and that whether it is increasing or decreasing is determined by a $Bernoulli(p=0.5)$ random variable.  $\mathbf{Y}$ is also either increasing or decreasing in $i$, as determined by an independent $Bernoulli(p=0.5)$ random variable. 
Each realization of $\mathbf{X,Y}$ manifests either a positive or negative association between $X$ and $Y$, and no single realization will ever give a null measure of association or correlation.  Specifically, with probability $0.5$ one of $\mathbf{X}$ and $\mathbf{Y}$ is increasing and the other decreasing, in which case they will be negatively associated, and with probability $0.5$ they are both either increasing or decreasing, and are therefore positively associated. Without additional assumptions, it is impossible to tease apart the association between $X$ and $Y$ from the shared dependence structure (in this case monotonicity in subject index) within $X$ and within $Y$.

This example highlights the fact that the parameter of interest--the $X-Y$ association--is a functional of the joint distribution $f_{\mathbf{X,Y}}$ of $\mathbf{X}$ and $\mathbf{Y}$, from which we have only a single draw. If we had access to data aggregated across draws from the joint distribution of $\mathbf{X,Y}$, e.g. $n$ draws from $f_{\mathbf{X,Y}}$, it would be easy to tell that $X$ and $Y$ are independent.  Similarly, if the data were i.i.d., then a single draw from $f_{\mathbf{X,Y}}$ would be equivalent to $n$ draws from $f_{X,Y}$ and there would be no problem.  But in the toy example, a single draw from $f_{\mathbf{X,Y}}$ does not suffice to identify the parameter of interest, in particular because there is no information in a single realization of $\mathbf{X,Y}$ about a crucial part of the data generating process, namely the Bernoulli coin flip that determines the direction of monotonicity.   

A hallmark of spurious associations due to dependence is that the distribution of the estimates of the association between $X$ and $Y$ is centered around the null value, even if no single-sample estimate ever attains it.  This distinguishes spurious associations from the consequences of  confounding, whereby an apparent association between $X$ and $Y$ may be due to an unmeasured third variable $Z$, e.g. if $Z$ is a common cause of $X$ and $Y$.  In general, having access to multiple draws from $f_{\mathbf{X,Y}}$  is not informative about bias due to unmeasured confounding, and estimates of the $X-Y$ association across multiple draws will be systematically biased.  On the other hand, having access to multiple draws from $f_{\mathbf{X,Y}}$ obviates the problem of spurious associations due to dependence.

Informally and heuristically, spurious associations due to dependence can arise whenever the similarity between $X$ values for two observations, e.g. $|X_i-X_j|$, predicts $Cov(X_i,X_j)$, which in turn predicts $Cov(Y_i,Y_j)$ and $|Y_i-Y_j|$.  The crucial step is the relationship between $Cov(X_i,X_j)$ and $Cov(Y_i,Y_j)$; this holds when  $\bf{X}$ and $\bf{Y}$ have similar variance-covariances matrices.  Therefore, $X$ can appear to carry information about $Y$, even if $X$ and $Y$ are in fact statistically independent of each other.  This can occur when $\mathbf{X}$ and $\mathbf{Y}$ exhibit temporal or genetic dependence, as has been long acknowledged in the literature, but also when they exhibit spatial, network, or any other kind of dependence that is shared (or similar) between $\mathbf{X}$ and $\mathbf{Y}$.  It will generally \emph{not} arise, however, whenever data are clustered, with dependence within but not between clusters.  This is because the hallmark of spurious associations due to dependence, namely that there is no systematic bias across draws from the data generating distribution, ensures that the spurious associations will tend to average out to the null value across independent clusters.
More formally, spurious associations due to dependence arise because we mistake a question about the joint distribution $f_{\mathbf{X,Y}}$ of the random vector $\mathbf{X}$ and the random vector $\mathbf{Y}$ for a question about the joint distribution $f_{X,Y}$ of $X$ and $Y$.  We treat a single draw from $f_{\mathbf{X,Y}}$ as if it were $n$ draws from $f_{X,Y}$.  

As the successes in time series and GWAS contexts demonstrate, there is a middle ground between the pathological toy example above, in which a single draw from $f_{\mathbf{X,Y}}$ contains literally no useful information about the association between $\mathbf{X}$ and $\mathbf{Y}$, and the i.i.d. setting, in which a single draw from $f_{\mathbf{X,Y}}$ is in fact equivalent to $n$ draws from $f_{X,Y}$, and suffices for statistical inference about the association between $X$ and $Y$.  In more realistic settings, such as the social network context we consider in this paper, we expect estimates of association to dispersed around the null value rather than concentrating on positive and negative values, with the level of dispersion controlled by the strength of the dependence within $\mathbf{X}$ and $\mathbf{Y}$ and the similarity of the dependence structures of  $\mathbf{X}$ and $\mathbf{Y}$.  We illustrate this in simulations in Section \ref{sec:dependence}; see Figure \ref{fig:correlation}.  

Although the problem of spurious associations due to dependence is most salient when $\mathbf{X}$ is independent of $\mathbf{Y}$, i.e. under the null, similar dependence structure between $\mathbf{X}$ and $\mathbf{Y}$ also results in spurious estimates when  $\mathbf{X}$ and $\mathbf{Y}$ are truly associated, i.e. under the alternative. In this case, because the direction of error due to shared dependence is random, the association between $X$ and $Y$ may be overestimated (when the direction of error due to spurious associations is in the same direction as the true association) or underestimated (when the direction of error due to spurious associations is in the opposite direction from the true association).

\section{Motivating Example: Framingham Heart Study}\label{sec:example}

The Framingham Heart Study (FHS), initiated in 1948, is arguably the most
important source of data on cardiovascular epidemiology. It is also an influential source of data on network peer effects. FHS is an ongoing cohort
study of participants from the town of Framingham, Massachusetts, that has grown over the years to include five cohorts with a total sample of over $15,000$. Study participants are followed through exams every 2 to 8 years. In between exams, participants are regularly monitored
through phone calls. Detailed information on data collected in
the FHS can be found in~\cite{tsao2015cohort}. Public versions
of FHS data through 2008 are available from the dbGaP database.  

Over 3000 papers using FHS data have been published in top medical journals, including influential findings such as effects of cholesterol, blood pressure, smoking, physical activity, and obesity on risk of heart disease \citep{dawber1957, dawber1959, kannel1967habitual, kannel1967relation}; effects of blood pressure, sleep apnea, and parental history on risk of stroke \citep{kannel1970epidemiologic, redline2010obstructive, seshadri2010parental}; identification of risk factors for dementia and Alzheimer's disease  \citep{schaefer2006plasma, au2006association, akomolafe2006diabetes, jefferson2015low}; and genetic markers for blood pressure, hypertension, heart disease, Alzheimer's disease, brain structure, and many other outcomes \citep{levy2009genome, seshadri2010genome, hibar2015common}.  The Framingham Risk Score, a simple algorithm for calculating 10-year risk of coronary heart disease based on FHS data, is commonly used for treatment decisions in clinical settings \citep{wilson1998prediction}. Because it is an ongoing study, hundreds of papers using these data continue to be published each year: Google Scholar lists 450 papers with "Framingham Heart Study" in the abstract published in the last 12 months.

In addition to its outsized role in cardiovascular disease epidemiology, the FHS plays a uniquely influential part in the study of social networks and peer effects (sometimes called "peer influence" or "social
contagion"). In the early 2000s, researchers Christakis and Fowler discovered an untapped resource buried in the FHS data collection
tracking sheets: information on social ties that, combined with existing data on connections among the FHS participants, allowed them to reconstruct
the (partial) social network underlying the cohort.  They then
leveraged this social network data to study peer effects for obesity~\citep{christakis2007spread},
smoking~\citep{christakis2008collective}, and happiness~\citep{fowler2008dynamic}.  Researchers have since used the same methods as Christakis and Fowler to study peer effects in the FHS and in many other social network settings (e.g.  \citealp{trogdon2008peer,fowler2008dynamic,rosenquist2010spread}).   

As is standard practice for cohort studies, publications using FHS data report statistical models that assume independent subjects (with the exception of some studies that use standard methods to account for family structure in the genetic data that was collected as part of FHS's later waves). This is despite the facts that the study population comprises close to a quarter of the total population of Framingham, MA, that more than $1,500$ extended families are represented by multiple members ($>3$) in the study population, and that many of the exposures and outcomes being studied have social or familial determinants that make them prime candidates for social network dependence.  In fact, even in the literature on peer effects, where the very hypotheses of interest imply non-independent subjects, researchers have almost exclusively relied on models, like generalized estimating equations, that assume independent subjects (while accounting for repeated measurements within subject).  In Section \ref{sec:FHS} we reanalyze several papers from this body of literature and provide evidence that network dependence undermines the assumptions on which the original analyses rest, potentially resulting in spurious association estimates, underestimated standard errors, and anti-conservative p-values.



\section{Network dependence}
\label{sec:dependence}

A network is a collection of nodes and edges, where, in a social network, a node
represents a person and an edge connecting two nodes represents the existence
of some relationship or social tie between them. In the FHS data, edges represent relationships like being genetically related, being married, and being neighbors. The adjacency matrix $\mathbf{A}$ for an $n$-node network is an $n \times n$ matrix with entries $A_{ij}$ indicating the presence and attributes of an edge between nodes $i$ and
$j$.  In this paper we consider binary symmetric adjacency matrices, representing simple undirected networks. However, the ideas that we present apply equally to directed networks, in which the presence of an edge from node $i$ to node $j$ does not imply an edge from node $j$ to node $i$, and to networks with different kinds or strengths of edges. Distance in a network is usually measured by \emph{geodesic distance}, a count of the number of edges along the shortest path between two nodes. 

When people are connected by social network ties, their data may be dependent, but this dependence is routinely ignored in both the methodological and applied literature using social network data.  In some settings researchers routinely account for statistical dependence in data analyses, for example, when data are clustered (e.g. clustered randomized trials, batch effects in lab experiments), when studying genetics or heritability in a sample of genetically related organisms, or when data may exhibit spatial or temporal dependence.  But outside of these settings it is generally standard practice to use statistical methods that assume i.i.d. data, and network dependence has previously received almost no attention in the statistical methods literature, let alone in the applied literatures in which it most often occurs. 

In social networks, we propose that network dependence can come from one or both of two sources.  \emph{Latent variable dependence} is due to latent traits that are more similar for observations that are close
than for distant observations. Homophily, or the tendency of similar people to form
network ties, is a paradigmatic source of latent variable dependence in social networks.
If the outcome under study in a social network has a genetic component,
then we would expect latent variable dependence due the fact that
family members, who share latent genetic traits, are more likely to
be close in social distance than people who are unrelated. If the
outcome is affected by geography or physical environment, latent variable
dependence could arise because people who live close to one another
are more likely to be friends than those who are geographically distant. The second source of dependence is \emph{direct transmission}: in networks,
edges often present opportunities to transmit traits or information
from one node to another, resulting in dependence that is informed by the underlying network structure. This type of dependence could affect behavioral or infectious outcomes.
In general, these sources of dependence result in positive
pairwise correlations that tend to be larger for pairs of observations
from nodes that are close in the network and smaller for observations
from nodes that are distant in the network \citep{ogburn2017challenges}.  They result in dependence that is analogous to spatial or temporal dependence, with the key difference (discussed briefly in Section \ref{sec:solutions}) that the underlying topology is likely to be highly non-Euclidean.  Network dependence is primarily a problem when data are collected from one (as in the FHS) or a very small number of interconnected networks. If data are collected from many independent networks, or equivalently if a network is comprised of many independent connected components, then it is straightforward to treat the independent (sub-)networks as independent clusters using existing methods.

To distinguish between the issue that a network sample may not be representative of a broader target population, which we ignore in this work, and the consequences of treating network observations as
if they are i.i.d. with which we are concerned, consider a hypothetical sample of $n$ nodes from the network underlying the FHS data.  Each node provides an outcome $Y$, e.g. body mass
index (BMI). Suppose that, as has been suggested by some researchers
\citep{christakis2007spread}, BMI exhibits network dependence due to social contagion. The target of inference is the mean $\mu$ of BMI for the U.S. population.  

Crucially, the bias of a sample mean is not 
necessarily affected by network dependence. The sample average $\bar{Y}=\sum\limits _{i=1}^{n}Y_{i}/n$ will have expectation equal to $\mu$ as long as the residents of Framingham do not systematically differ from the overall U.S.  population in terms of BMI. To put this another way, although it is not a random sample from the target population, a network-dependent sample may still be a representative draw from the true underlying distribution of $Y$.  (If this seems implausible, suppose instead that the target of inference were the mean BMI for the adult population of largely middle class, mid-sized cities in New England, of which Framingham is an average example.)
However, the variance of $\bar{Y}$ will generally be underestimated unless dependence is taken into account.  Specifically, $var(\bar{Y})=\sum_{i}var(Y_i)+\sum_{i\neq j}cov(Y_i,Y_j)$, but a researcher who ignores dependence will leave off the second term, which is expected to be strictly positive under latent variable dependence and dependence due to direct transmission.
The problem, then, is that finite sample estimation error due to sampling variability, which should be captured by a standard error estimate, confidence interval, or p-value, is not accurately reflected in an inferential procedure that assumes independent observations. With increasing dependence, inference that assumes independence tends to be increasingly anticonservative.  

If researchers are interested in learning about the association between an exposure, $X$, and an outcome, $Y$, using observations sampled from nodes in a social network, and if both $X$ and $Y$ exhibit network dependence, then spurious associations due to dependence can arise.  
Even if the goal of inference is purely predictive, spurious associations due to dependence undermine the ability of a model fit to a network-dependent sample to provide unbiased out-of-sample predictions, since the sign of the error in estimating the $X$-$Y$ association due to shared dependence is random.  There has been much discussion of the challenge that confounding due to homophily poses for learning about peer effects \citep{shalizi2011homophily}.  This usually refers to causal confounding; however, if a predictor and outcome both exhibit network dependence due to different latent variables, then homophily could be implicated in the shared dependence that gives rise to spurious associations. 

\begin{figure}[H]
	\captionsetup{format=plain}
	\centering
	\subcaptionbox{	\label{fig:corr_iid}Correlation between iid $X$ and iid $Y$}[0.40\linewidth]
	{\includegraphics[width=0.4\textwidth]{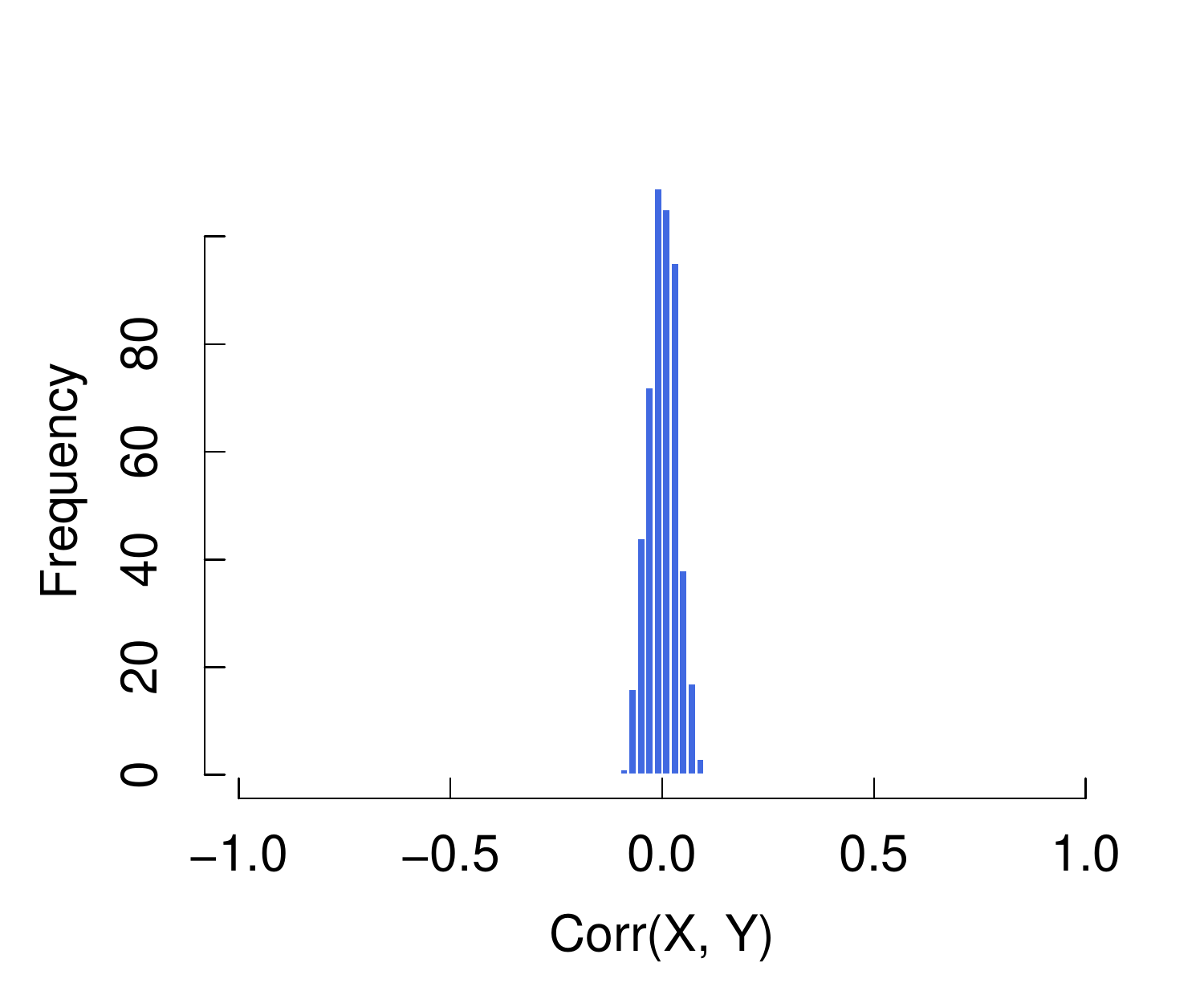}}
		\hspace*{0.5\fill}
	\subcaptionbox{	\label{fig:corr_large_e}
	Correlation between $X$ and $Y$ generated under direct transmission with large random errors}[0.40\linewidth]
			{\includegraphics[width=0.4\textwidth]{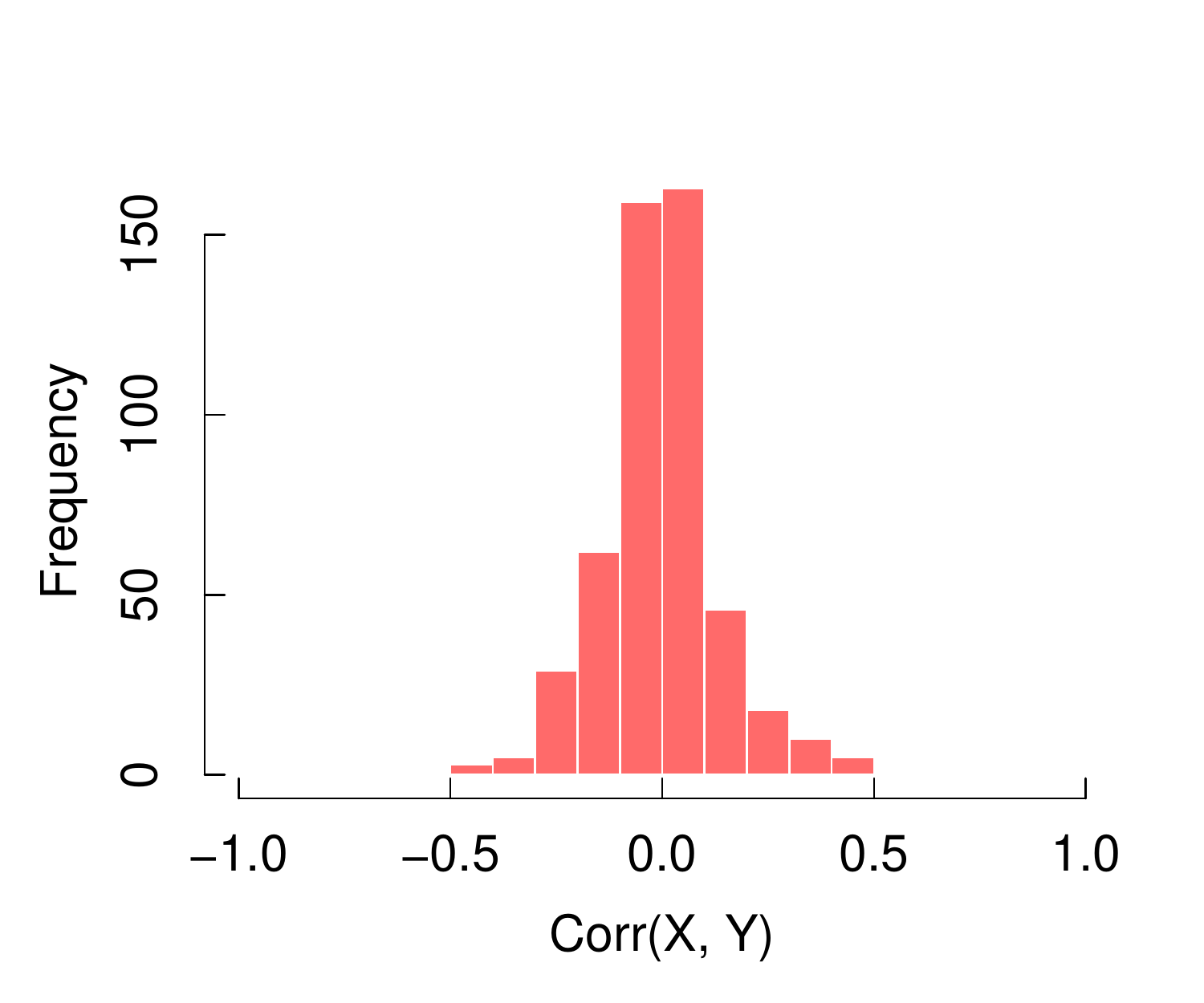}}
		
	\subcaptionbox{	\label{fig:corr_moderate_e}
		Correlation between $X$ and $Y$ generated under direct transmission with moderate random errors}[0.40\linewidth]
			{\includegraphics[width=0.4\textwidth]{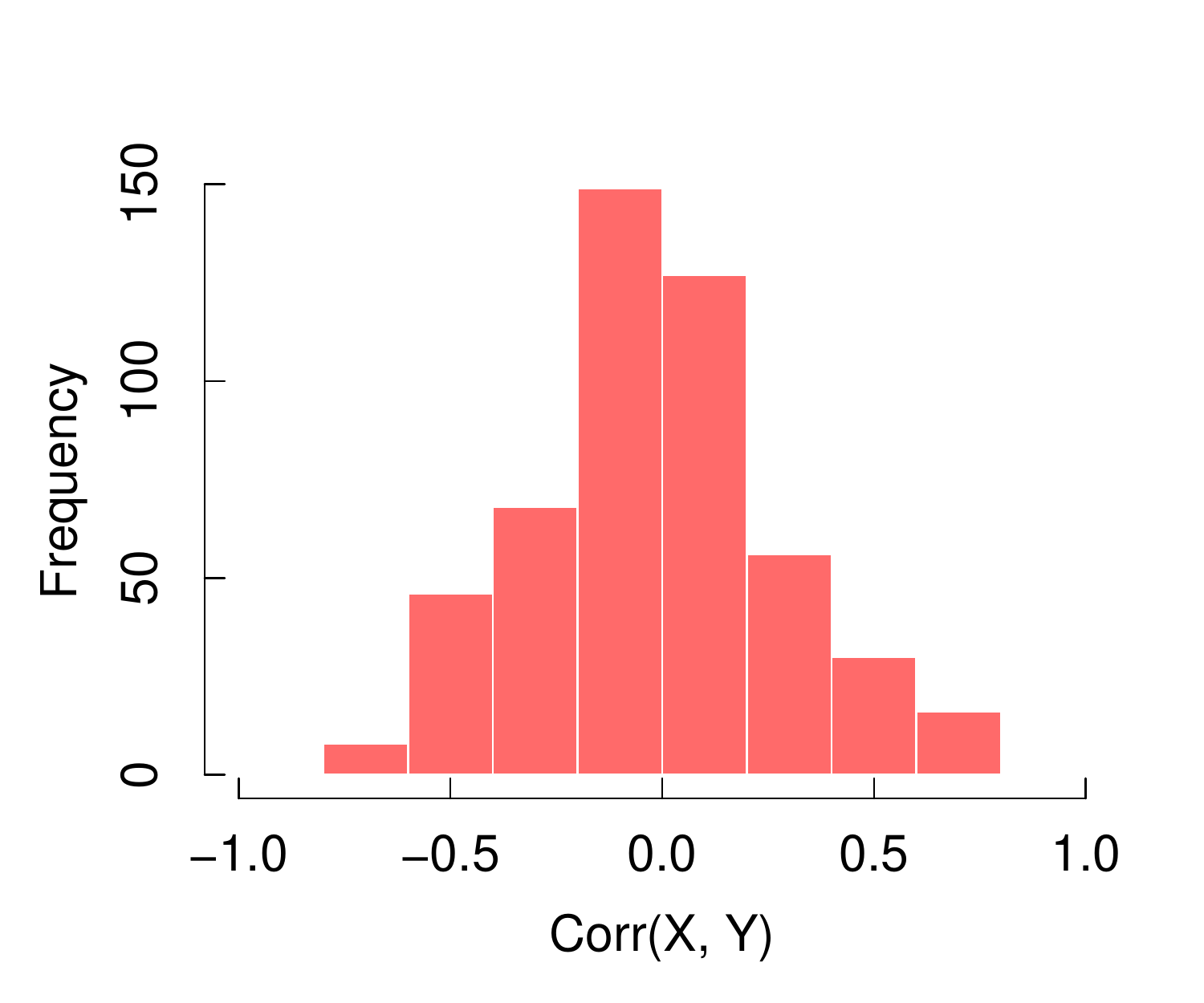}}
		\hspace*{0.5\fill}
		\subcaptionbox{	\label{fig:corr_small_e}
		Correlation between $X$ and $Y$ generated under direct transmission with small random errors}[0.40\linewidth]
			{\includegraphics[width=0.4\textwidth]{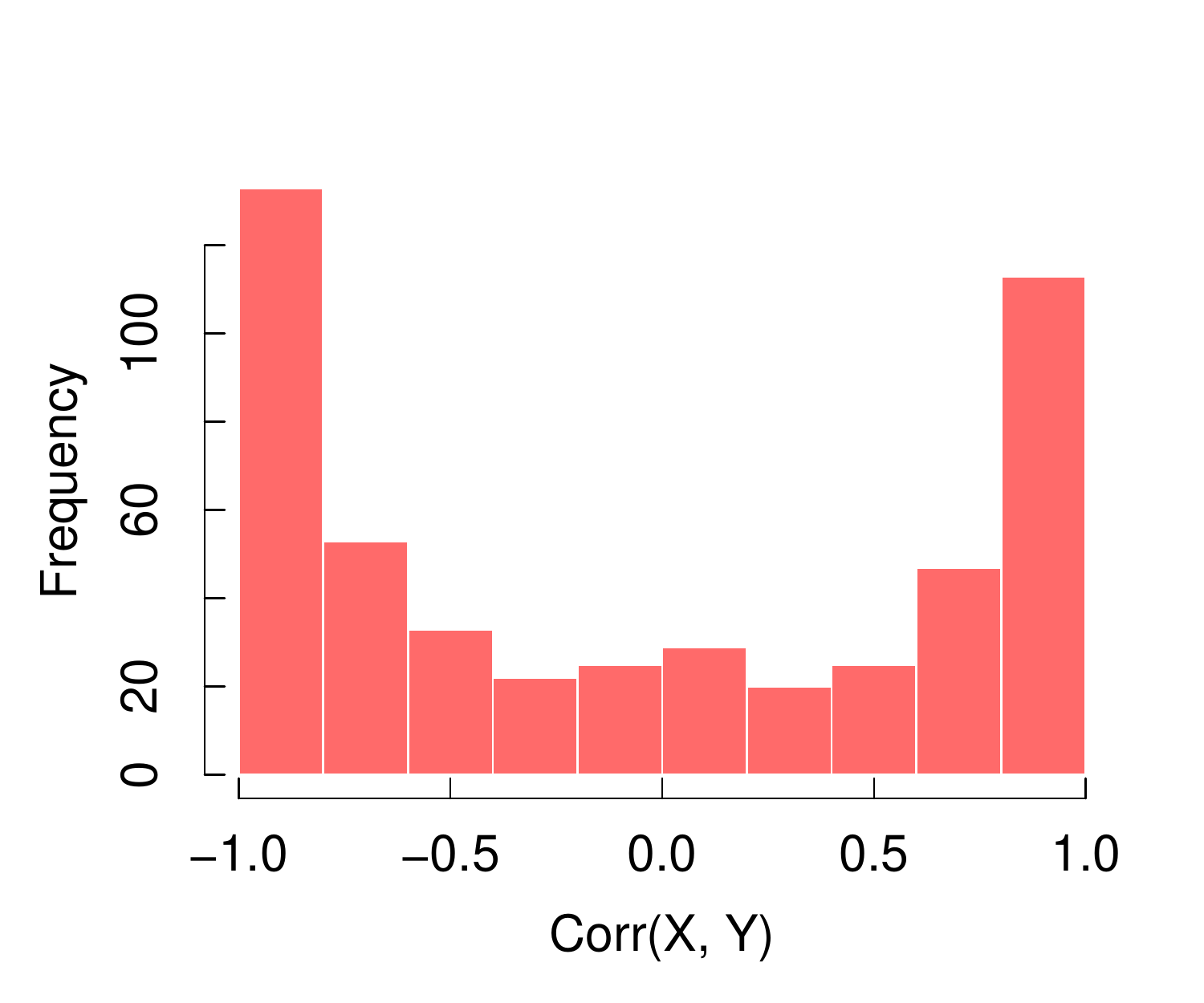}}
	\caption{}
	\label{fig:correlation}
\end{figure}

Depending on the nature of the dependence exhibited by $\bf{X}$ and $\bf{Y}$, spurious associations due to network dependence can result in a distribution of measures of associations that is similar to, but has greater variance than, the distribution expected when $\bf{X}$ and $\bf{Y}$ are i.i.d. In these cases it may be tempting to conclude that the problem is simply underestimated variance.  However, a hallmark of spurious associations is a clear pattern of inflated estimates, with a symmetric, bimodal distribution of measures of association concentrated away from $0$.  We illustrate this in simulations presented in figure \ref{fig:correlation}. We simulated $\bf{X}$ and $\bf{Y}$ under three different direct transmission processes, varying the relative contribution of influence from a node's neighbors and a unit-specific random error term at each step in the direct transmission process.  We also simulated a setting with no network dependence, in which $\bf{X}$ and $\bf{Y}$ are i.i.d. $N(0,1)$ random variables. In all four settings $\bf{X}$ is independent of $\bf{Y}$, but in the three network dependent settings $\bf{X}$ and $\bf{Y}$ are generated under the same direct transmission model, resulting in similar covariance structures due to the same underlying network. For each setting we ran 500 simulations and calculated the correlation between $\bf{X}$ and $\bf{Y}$ in each simulation.  The distribution of correlation coefficients is shown in Figure \ref{fig:corr_iid}. When the random error is large relative to the influence term, that is when dependence within $\bf{X}$ and within $\bf{Y}$ is weak, the distribution of correlation coefficients is similar to, but with greater variance than, the distribution for i.i.d. $\bf{X}$ and $\bf{Y}$.  However, as the error terms get smaller, that is as the dependence within $\bf{X}$ and within $\bf{Y}$ is strong, the distribution of correlation coefficients converges to a bimodal distribution concentrated around $-1$ and $1$, as shown in Figure \ref{fig:corr_small_e}.  

This pattern is different from what we would see under causal confounding, where the correlation between $X$ and $Y$ would tend to be systematically positive or negative.  It is also different from what we would see if the problem were merely that treating $\bf{X}$ and $\bf{Y}$ as i.i.d. resulted in underestimating the variance of correlation/association measures.  That story is consistent with Figures \ref{fig:corr_large_e} and \ref{fig:corr_moderate_e}, but not with Figure \ref{fig:corr_small_e}.

While the solution to causal confounding is, in principle, simple--control for confounders--the solution to dealing with spurious associations due to dependence must be tailored to the unique features shared dependence structures in different settings.  The solutions that researchers found for dealing with confounding by cryptic relatedness in GWAS data, and with spurious associations in econometric time series data, are each highly specific to known features of the data-generating processes in those two settings.  We do not have easy solutions to propose for dealing with spurious associations due to dependence in general, or in social networks, since it is difficult to imagine that a one-size-fits-all approach could encompass all of the ways that social networks can exhibit network dependence.  We briefly discuss future research directions to this end in Section \ref{sec:solutions}.

The two issues of anticonservative standard error estimation and spurious associations due to network dependence converge in  standard regression analyses, which are used throughout the applied literature on social networks.  These models assume independent errors, but when data exhibit network dependence the regression errors may, too, undermining inferences drawn from the regression models.  Furthermore, as demonstrated in Section 12.2.3 of \cite{kass2014analysis} in the context of time series, regression coefficients are plagued by spurious associations due to dependence.  Although researchers have developed regression models for many kinds of dependent data, it is not clear that any of them are generally appropriate for social network data, and certainly none are in wide use for network data.  In Sections \ref{subsec:epi} and \ref{subsec:peer} we find evidence of network dependence in the regression residuals from models published with FHS data. Though we cannot say whether any substantive conclusions are true or false, we can find (or fail to find) evidence that researchers used inferential procedures that rest on false assumptions; this is what we refer to as \emph{invalid inference}.

In Section \ref{sec:test} we propose a way to test for the possibility of spurious associations due to network dependence, in Section \ref{sec:sims} we illustrate the phenomenon of spurious associations in simulations, and in Sections \ref{subsec:epi} and \ref{subsec:peer} we find evidence of possible spurious associations due to network dependence in published papers using FHS data.

\section{Testing for network dependence}\label{sec:test}

In this section we describe a test of the null hypothesis of no network dependence.  More precisely, the null hypothesis is of no systematic relationship between network topology and the values of the nodes' random variables, where what constitutes "systematic" can be controlled by a choice of weight matrix.  This null hypothesis space includes joint independence of the random variables, and by our choice of weight matrix we can ensure that no form of network dependence (that is, dependence that decays with network distance) is included in the null hypothesis space \citep{assuncao1999new}. We will therefore use "independence" as a shorthand for this null hypothesis.

Moran's $I$ is a popular test for spatial autocorrelation that is known to work well whenever data are distributed according to a simultaneous autoregression (SAR) model \citep{black1992network, butts2008social, long2015social, fouss2016algorithms}.  
However, we propose that Moran's $I$ can in fact be used to test for any kind of network dependence that is positive and inversely related to network distance. Moran's $I$ provides a valid and unbiased test for network dependence--that is, it has the expected null distribution under independence and that it has non-null power under the alternative of network dependence.  We verified in simulations that the power of Moran's $I$ to detect dependence tends to increase with increasing dependence (see Section \ref{sec:sims}).

In spatial settings, Moran's $I$ takes as input an $n$-vector of continuous random variables and an $n\times n$ weighted distance matrix $\mathbf{W}$, where entry
$w_{ij}$ is a non-negative, non-increasing function of the Euclidean
distance between observations $i$ and $j$. Moran's $I$ is expected
to be large when pairs of observations with greater $w$ values (i.e.
closer in space) have larger correlations than observations with smaller
$w$ values (i.e. farther in space). The choice of non-increasing
function used to construct $\mathbf{W}$ is informed by background
knowledge about how dependence decays with distance; it affects the
power but not the validity of tests of independence based on Moran's
$I$. Geary's $c$
\citep{geary1954contiguity} is another statistic commonly used to
test for spatial autocorrelation~\citep{fortin1989spatial,lam2002evaluation,da2008diagnosis};
it is very similar to Moran's $I$ but more sensitive to local, rather
than global, dependence. We focus on Moran's $I$ in what follows
because our interest is in global, rather than local, dependence.
Because of the similarities between the two statistics, Geary's $c$
can be adapted to network settings much as we adapt Moran's $I$.

Let $Y$ be a continuous variable of interest and $y_{i}$ be its realized observation
for each of $n$ units~$(i=1,2,\ldots,n)$. Each observation is associated with a location, traditionally in space but we will extend this to networks.  
Then Moran's $I$ is defined as follows:
\begin{eqnarray}
I=\frac{\sum\limits _{i=1}^{n}\sum\limits _{j=1}^{n}w_{ij}\big(y_{i}-\bar{y}\big)\big(y_{j}-\bar{y}\big)}{S_{0}\sum\limits _{i=1}^{n}\big(y_{i}-\bar{y}\big)^{2}/n},\label{eq:moran}
\end{eqnarray}
where $S_{0}=\sum\limits _{i=1}^{n}\sum\limits _{j=1}^{n}(w_{ij}+w_{ji})/2$ and $\bar{y}=\sum\limits _{i=1}^{n}y_{i}/n$.
Under independence, the pairwise products $(y_{i}-\bar{y})(y_{j}-\bar{y})$
are each expected to be close to zero. On the other hand, under network dependence close
pairs are more likely to have similar values than distant pairs, and
$(y_{i}-\bar{y})(y_{j}-\bar{y})$ will tend to be relatively large
for the upweighted close pairs; therefore, Moran's $I$ is expected
to be larger in the presence of network dependence than under the
null hypothesis of independence.

Tests for spatial dependence take Euclidean distances (usually in $\mathbb{R}^{2}$
or $\mathbb{R}^{3}$) as inputs into the weight matrix $\mathbf{W}$. In networks, the entries in {$\mathbf{W}$} can be
comprised of any non-increasing function of geodesic distance, but for robustness in what follows we use the adjacency matrix $\mathbf{A}$, where $A_{ij}$ is an indicator of nodes $i$ and
$j$ sharing a tie. The choice of $\mathbf{W}=\mathbf{A}$ puts weight
1 on pairs of observations at a distance of $1$ and weight $0$ otherwise.
In many spatial settings, subject matter expertise can facilitate
informed choices of weights for $\mathbf{W}$ (e.g.~\citealt{smouse1999spatial,overmars2003spatial}),
but it is harder to imagine settings where researchers have information
about how dependence decays with geodesic network distance. Dependence due to direct transmission is transitive: dependence between
two nodes at a distance of $2$ is through their mutual contact. This
kind of dependence would be related to the number, and not just length,
of paths between two nodes. It may also be possible to construct distance
metrics that incorporate information about the number and length of
paths between two nodes, but this is beyond the scope of this paper.
In general, in the presence of network dependence adjacent nodes have
the greatest expected correlations; therefore $\mathbf{W}=\mathbf{A}$
is a valid choice in all settings. Of course, if we have knowledge
of the true dependence mechanism, using a weight matrix that incorporate
this information will increase power.

The standardized statistic $I_{std}:=(I-\mu_{I})/\sqrt{\sigma_{I}^{2}}$
is asymptotically normally distributed under mild conditions on $\mathbf{W}$ and $Y$~\citep{sen1976large}; using the known asymptotic distribution of the test statistic under the null permits hypothesis tests of independence using the normal
approximation. However, for network data we propose to run permutation tests by
permuting the $Y$ values associated with each node while holding
the network topology constant. This tests the null hypothesis of no systematic relationship between pairs of random variables $(Y_i,Y_j)$ and the corresponding distances between nodes $i$ and $j$. Setting $w_{ij}=0$ for all non-adjacent
pairs of nodes results in increased variability of $I$ relative to
spatial data, and therefore the normal approximation may require larger
sample sizes to be valid for network data compared to spatial data.
This permutation test is valid regardless of the distributions of $\mathbf{W}$
and $Y$ and for small sample sizes. In the Appendix we formalize the permutation algorithm and show that the mean and variance of the test statistic under the permutation distribution correspond to the expected null distribution moments. 

In a companion paper 
\citep{leeogburn2018}, 
we propose a new test for spatial or network dependence in categorical random variables, which are common in social network settings (e.g. group membership). An R package for both tests of network dependence is available 
\citep{lee2018netdep}. 

We recommend viewing moderate to large statistics as evidence of possible dependence even if $p$-values do not meet an arbitrary $\alpha=0.05$ cut-off, and caution that network dependence may be present even if these statistics are small. Evidence based on Moran's $I$ does not directly speak to the accuracy of the substantive conclusions of any analysis; it can only provide evidence against the validity of the independence assumption on which an analysis relies. 
If the test statistic calculated from regression residuals is moderate to large, it suggests that  standard error estimates from i.i.d. regression models may be underestimated. If both of the test statistics calculated from an outcome and from a covariate of interest are moderate to large, it suggests that estimates of association may be spurious due to network dependence.

\section{Simulations}
\label{sec:sims}

In order to demonstrate that Moran's $I$ provides valid tests for network dependence, we simulated random variables associated with nodes in a single interconnected network, with dependence structure informed by the network ties. For each of four simulation settings we generated a fully connected social network with $n=200$ nodes.  We simulated i.i.d., mean-zero starting  values for each node and then ran several iterations of a direct transmission process, by which each node is influenced by its neighbors, to generate a vector of dependent outcomes $\mathbf{Y}=(Y_{1},Y_{2},...,Y_{200})$ associated with the nodes.  We ran the simulation $500$ times for each setting, generating $500$ outcome vectors.  While the amount of network dependence in the outcomes varied across simulation settings (controlled by the number of iterations of the spreading process), the expected outcome $E[Y]$ was $0$ for every setting.  To demonstrate the impact of using i.i.d. methods when dependence
is present, in each simulation we calculated a $95\%$ confidence interval (CI) for $E[Y]$ under
the assumption of independence.  The CI is given by $\bar{Y}\pm1.96*s.e.$, where we estimated the standard error (s.e.) for 
$\bar{Y}$ under the assumption of independence, that is ignoring the presence of any pairwise covariance terms. In each simulation we also ran a test for network dependence using Moran's $I$.  

\begin{figure}[ht]
	\centering
	\includegraphics[width=\linewidth]{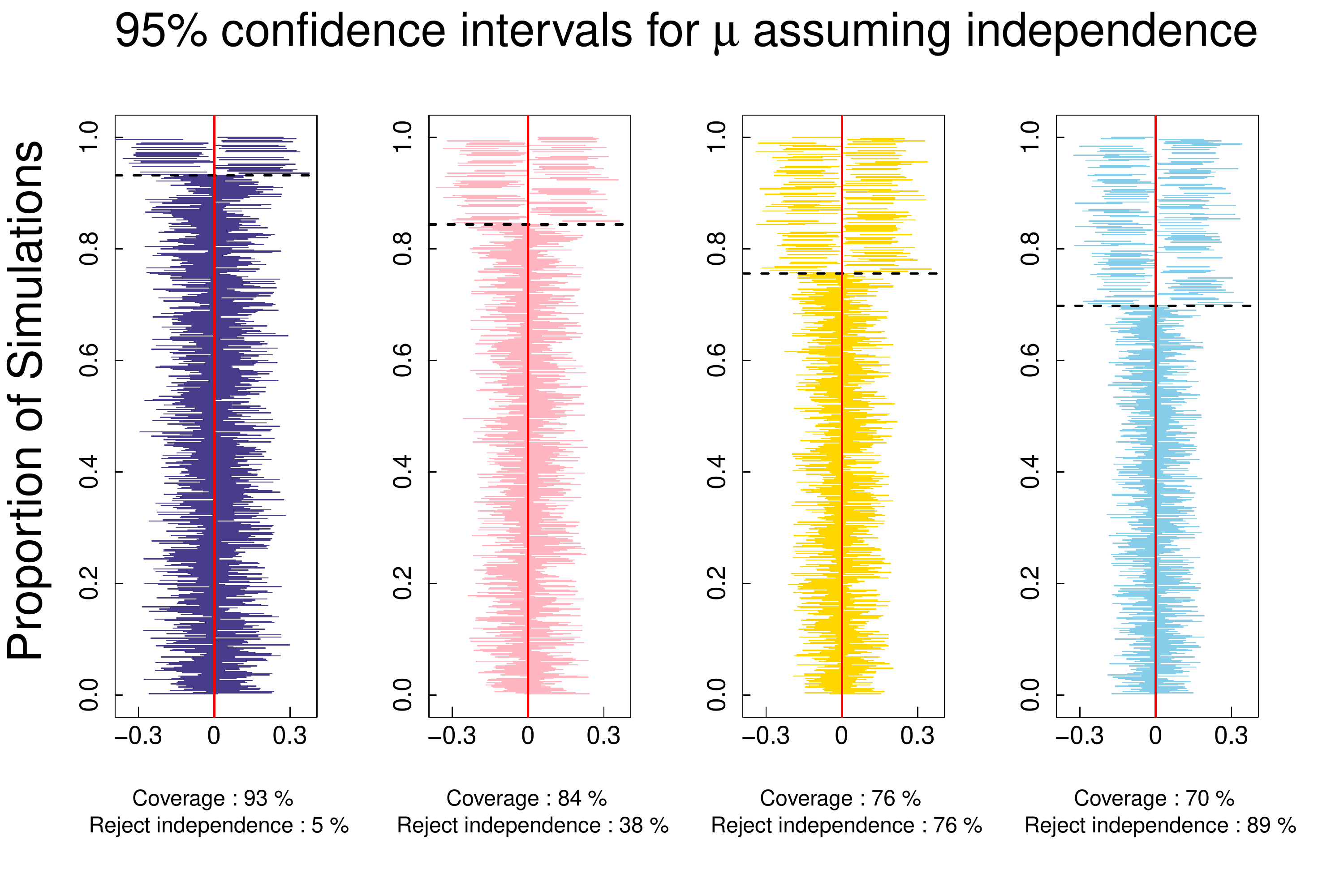}
	\caption{Each column contains 95$\%$ confidence intervals (CIs) for $E[Y]=\mu$
		under dependence due to direct transmission, with increasing dependence from left (no dependence) to right. The CIs above the dotted
		line do not contain the true $\mu=0$ (red-line) while the CIs below
		the dotted line contain $\mu$. Coverage rates of 95$\%$ CIs are
		calculated as the percentages of the CIs covering $\mu$. We also present the percentages of permutation tests based on Moran's $I$ that reject the null at $\alpha=0.05$; this is the type I error for the leftmost column and the power for the other three columns.}
	\label{fig:coverage_peer}
\end{figure}

\begin{table}[ht]
\centering
\caption{\label{tab:bias_coverage} Bias, average of the absolute values of the estimation errors, estimated standard errors, and true standard deviations corresponding to the columns in Figure \ref{fig:coverage_peer}.}
\resizebox{0.7\textwidth}{!}{\begin{tabular}{rllll}
  \hline
 column & 1 & 2 & 3 & 4 \\ 
  \hline
bias & 0.002 & 0.001 & 0.001 & 0.001 \\ 
 mean $|$estimation error$|$ & 0.059 & 0.066 & 0.068 & 0.071 \\ 
  mean estimated standard error & 0.071 & 0.059 & 0.051 & 0.046 \\ 
  standard deviation of the estimates & 0.076 & 0.083 & 0.086 & 0.090 \\ 
   \hline
\end{tabular}}
\end{table}

Figure~\ref{fig:coverage_peer} displays the results of four simulation settings, with increasing dependence from left to right. The left-most column represents a setting with no dependence.  Each column depicts $500$ $95\%$ confidence intervals, one for each simulation. The confidence intervals below the dotted lines cover the true mean of $0$, while the intervals above the dotted line do not. The coverage is close to the nominal $95\%$ under independence, but decreases dramatically as dependence increases, despite the fact that $\bar{Y}$ remains unbiased for $E[Y]$.  The diminished coverage is due in part to decreasing estimated standard errors, as reported in Table \ref{tab:bias_coverage}, and the resulting narrowing confidence intervals.

We report the power of permutation tests based on Moran's $I$ (with subject index randomly permuted $M=500$ times) to reject the null hypothesis of independence at the $\alpha=0.05$ level.  Under independence the test rejects 5\% of the time, as is to be expected, and as dependence increases and coverage decreases, the power of our test to detect dependence increases, achieving almost 90\% when the coverage drops below 70\%.  That the power to detect dependence increases with increasing dependence is robust to the specifics of the simulations, but the exact relation between coverage and power is not; in other settings 90\% power could correspond to different coverage rates, highlighting the fact that a strict  $p<0.05$ cut-off may not be appropriate for these tests of dependence.  Details are available in the Appendix, along with analogous results from additional simulations with latent variable dependence.  

In order to illustrate spurious associations due to network dependence, we simulated pairs a predictor, $X$, and outcome, $Y$, for each node, so that $X$ is independent of $Y$ but $\bf{X}$ and $\bf{Y}$ both have network dependence related to the same underlying network structure. This time we simulated three settings, with increasing dependence in both $\bf{X}$ and $\bf{Y}$ (indexed by $\kappa$).  For the setting without spurious associations, we permuted the $Y$ values from the setting with $\kappa=3$.  For each of the four settings, in each of 500 simulated datasets we regressed $Y$ on $X$ and an intercept and calculated a 95\% confidence interval for the $X$ coefficient, $\beta$.  

\begin{figure}[ht]
	\centering
	\includegraphics[width=\linewidth]{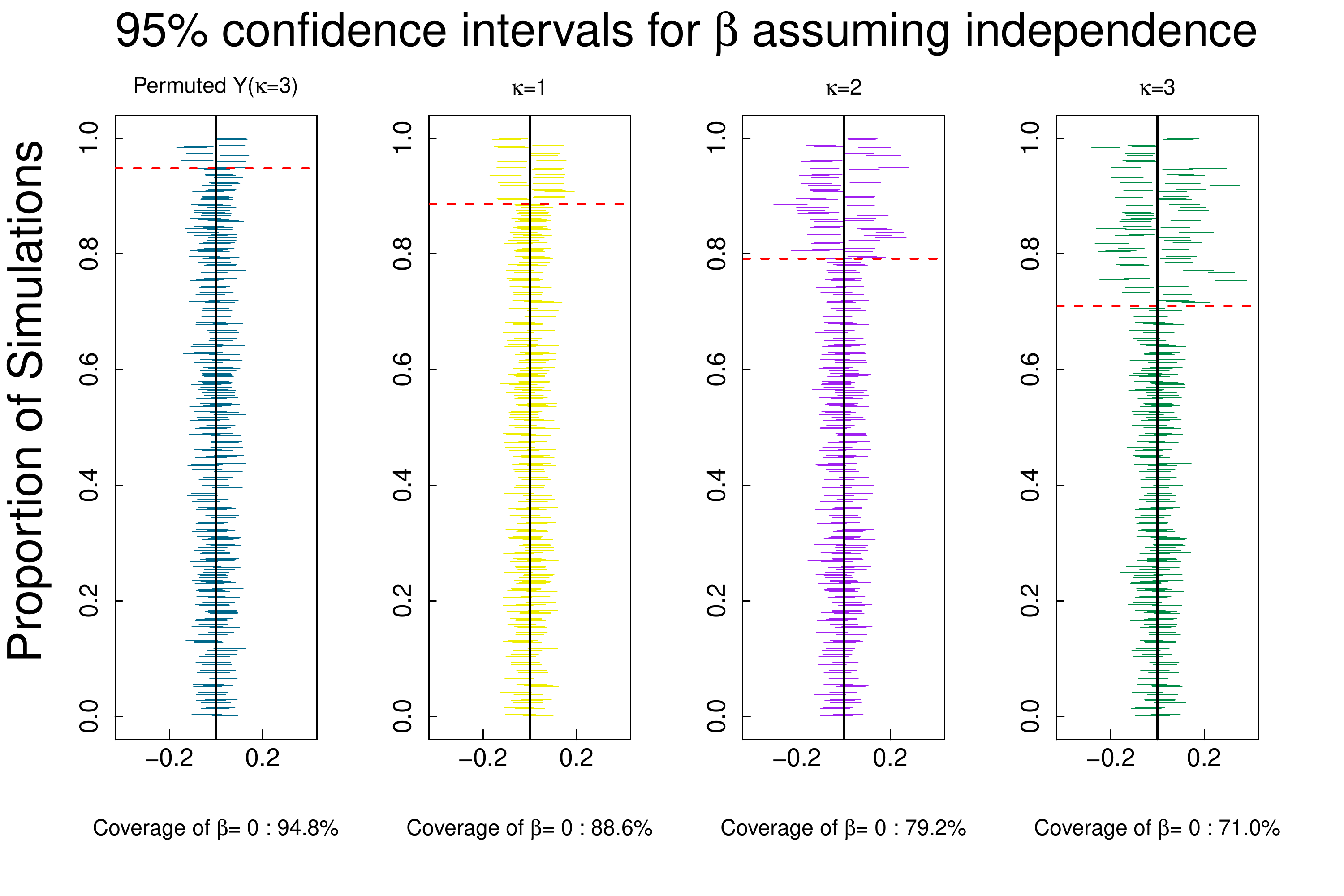}
	\caption{Each column contains 95$\%$ confidence intervals (CIs) for the $X$ coefficient, $\beta$, from a regression of $Y$ on 	
		$X$, estimated as if the data were i.i.d. 
		$X$ and $Y$ were simulated with increasing dependence from left (no dependence) to right. The CIs above the dotted
		line do not contain the true $\beta=0$ (red-line) while the CIs below
		the dotted line contain $0$. Coverage rates of 95$\%$ CIs are
		calculated as the percentages of the CIs covering $0$.}
	\label{fig:confounding}
\end{figure}

Figure~\ref{fig:confounding} displays the results of four simulation settings, with increasing dependence from left to right.  Each column depicts $500$ $95\%$ confidence intervals, one for each simulation. The confidence intervals below the dotted lines cover the true value $0$, while the intervals above the dotted line do not. The coverage is close to the nominal $95\%$ in the left-most column, but decreases dramatically as dependence increases.  In contrast to the simulations in Figure~\ref{fig:coverage_peer}, the width of the confidence intervals stays the same as dependence increases, but the average estimation error of $\hat{\beta}$ increases as dependence increases. Table \ref{tab:bias_confounding} shows that the estimated standard error remains constant while the absolute value of the estimation errors--and therefore true standard deviation--increases.

\begin{table}[ht]
\centering
\caption{\label{tab:bias_confounding} Bias, average of the absolute values of the estimation errors, estimated standard errors, and true standard deviations corresponding to the columns in Figure \ref{fig:confounding}.}

\resizebox{0.7\textwidth}{!}{\begin{tabular}{rrrrr}
  \hline
 & permuted $Y$ & $\kappa$ = 1 & $\kappa$ = 2 & $\kappa$ = 3 \\ 
  \hline
bias & -0.000 & 0.001 & 0.001 & 0.002 \\ 
 mean  $|$estimation error$|$ & 0.024 & 0.030 & 0.041 & 0.053 \\ 
  mean estimated standard error & 0.031 & 0.031 & 0.031 & 0.031 \\ 
  standard deviation of the estimates & 0.031 & 0.038 & 0.055 & 0.074 \\ 
  reject independence (Y) & 5\% & 80\% & 100\% & 100\% \\ 
  reject independence (X) & 100\% & 80\% & 100\% & 100\% \\ 
  reject independence (residuals) & 4\% & 80\% & 100\% & 100\% \\ 
   \hline
\end{tabular}}
\end{table}

Spurious associations due to network dependence are a result of covariance structure that is shared by $\bf{X}$ and $\bf{Y}$.  Causal confounding, which results in systematic bias, may result if network structure affects the mean, in addition to the covariance, of a random variable. To illustrate this systematic bias, we simulated a covariate with dependence structure governed by the FHS social network but otherwise unrelated to any of the variables measured in the FHS.  We generated a continuous network-dependent covariate, $\bf{X}$, conditional on the FHS network, independently 500 times. The mean of $X$ was higher for highly connected nodes than for isolated nodes.
 We regressed a cardiovascular outcome (systolic blood pressure, SBP), a lifestyle outcome (employed or not), a health-seeking behavior outcome (visited a doctor due to illness), and a non-cardiovascular health outcome (diagnosis of corneal arcus) from the FHS data onto $\bf{X}$.  For each of the four outcomes we fit the same regression model independently 500 times, once for each independently generated covariate.  

\begin{figure}[!h]
	\centering
	\includegraphics[width=\linewidth]{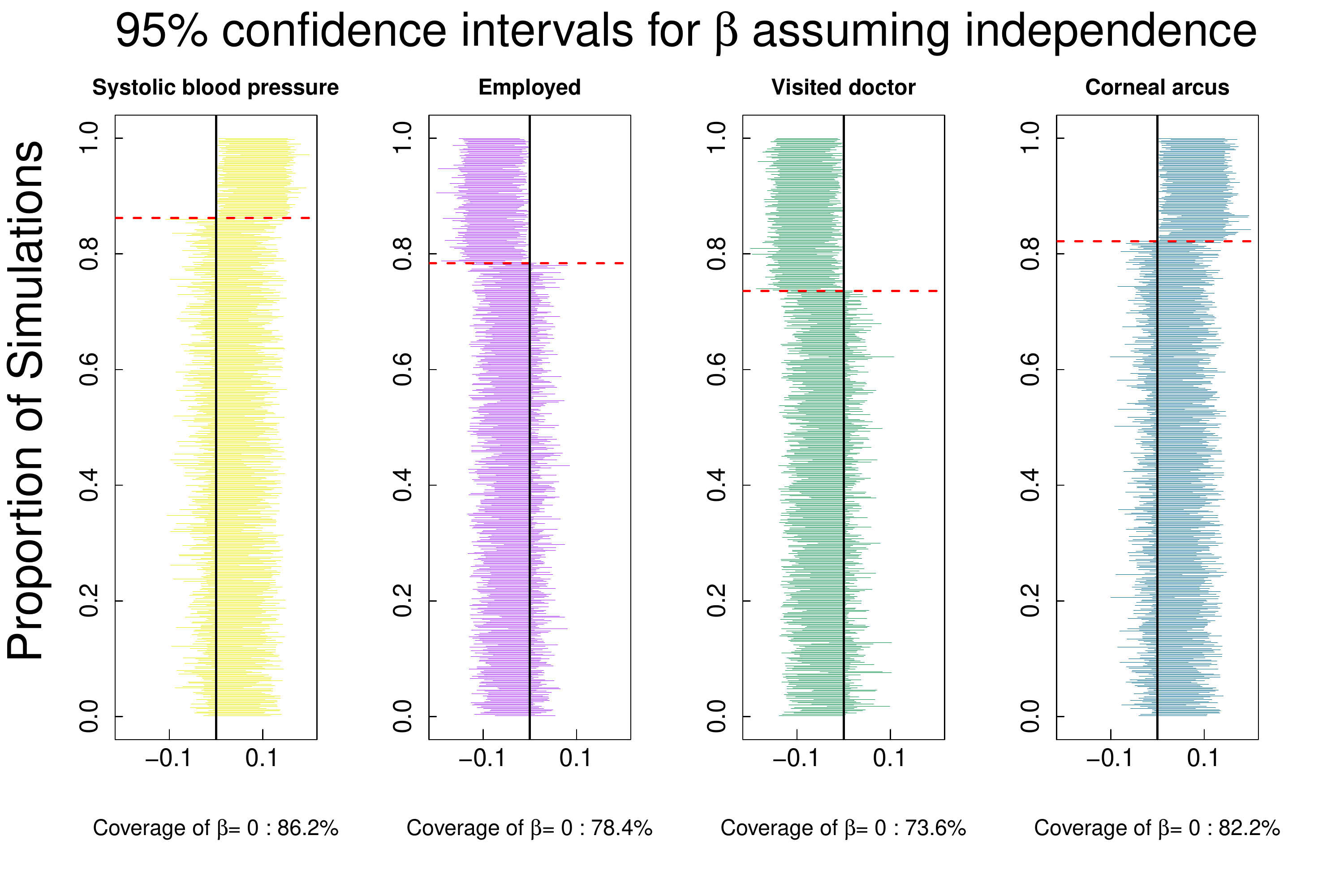}
	\caption{ {\bf 95\% confidence intervals under causal confounding by network connectedness.} Each column contains 95$\%$ confidence intervals (CIs) for the coefficient for a random, network dependent covariate. The CIs above the dotted line do not contain the null value $\beta=0$ (red-line) while the CIs below
		the dotted line contain $0$. Coverage rates of 95$\%$ CIs are
		calculated as the percentages of the CIs covering $0$.}
	\label{fig:fiveCIs}
\end{figure}

Figure \ref{fig:fiveCIs} shows the coverage of $95\%$ confidence intervals for $\beta$, the coefficient for $X$ in the regression of each outcome onto $X$ plus an intercept.  Because the covariate is generated without reference to any of these outcomes, the true value of $\beta$ for a population-based, rather than network, sample is plausibly $0$. In particular, due to the way it was generated, $X$ has no predictive value for any of these outcomes beyond the connectedness of the network node, which serves as a causal confounder for the relationship between $X$ and $Y$. However, for all four outcomes the confidence intervals are not centered around $0$. For all four outcomes the confidence intervals exhibit undercoverage, ranging from 74\% to 85\% rather than the nominal rate of 95\%.  The undercoverage may be due to both confounding and to network dependence in the regression residuals, which could result in underestimated standard errors.   Table \ref{tab:confounding} reports the $p$-values for tests of dependence in the four outcomes, the predictor $\bf{X}$ (averaged across 500 replicates), and the residuals from the regression of the outcome on $X$ (averaged across 500 replicates for each outcome). For three of the outcomes (SBP, employment, and corneal arcus) tests based on Moran's $I$ suggested strong evidence of dependence; for visit to doctor the test did not show strong evidence of dependence in the outcome or residuals (though we reiterate that a null test does not imply a lack of dependence).   Controlling for the confounder, namely nodal connectedness, recovers point estimates centered around $0$ (results in Appendix), but it may not fully address undercoverage due to network dependence.

Details for all simulations and analyses are in the Appendix.

\begin{table}[!ht]
	\centering
	\caption{\label{tab:confounding} Results of tests of network dependence for the outcomes, simulated predictor $X$, and residuals from regressing each outcome onto $X$. $P$-values are obtained from permutation tests.}
	\resizebox{1.0\textwidth}{!}{\begin{tabular}{|r|l|l|l|l|}
		\hline
		& Systolic blood pressure & Employed & Visited doctor & Corneal arcus \\ 
		\thickhline
		$p$-value for outcome & 0.03 & 0.00 & 0.71 & 0.01 \\ \hline
		Average $p$-value for predictor & 0.00 & 0.00 & 0.00 & 0.00 \\ \hline
		Average $p$-value for residuals & 0.02 & 0.00 & 0.71 & 0.10 \\ 
		\hline
	\end{tabular}}
\end{table}

\section{Analysis of Framingham Heart Study Data} \label{sec:FHS}

We found evidence of potentially widespread dependence in the outcomes, predictors, and regression residuals from published papers using FHS data.  The problem of network dependence extends to high profile research using FHS data to explicitly study peer effects and social contagion in social networks, but with statistical methods designed for i.i.d. data.

\subsection{Cardiovascular disease epidemiology} \label{subsec:epi}

In order to evaluate whether network dependence and spurious associations due to network dependence may undermine research using FHS data, we chose regression models from five published papers in the epidemiologic and medical literature and applied our tests of dependence to the outcomes, covariates, and regression residuals. We screened for ease of replicability using publicly available data (i.e. models are explicitly defined using variables that are available in the public data), and selected the first five papers that we found on Google Scholar that met the replicability criteria. Because we require social network information for our tests of dependence, and because that information is not available for all individuals and is not straightforward to harmonize across exams, we ran the published regression models on subsets of the data for which network information was readily available. Below we report results from the two papers for which we found the strongest evidence of dependence: the models reported in these two papers show compelling evidence of network dependent outcomes, covariates, and residuals. We also found moderate evidence of dependence in some of the analyses reported in each of the other three papers  \citep{wolf1991atrial, gordon1977high,levy1990prognostic}; details are in the Appendix. 

\cite{lauer1991impact} examined the association between obesity and
left ventricular mass (LVM); this
paper is one of the authors' many highly cited papers on LVM, which is of interest to many researchers due to its relationship with cardiovascular disease~\citep{levy1990prognostic} and other cardiovascular outcomes. The study assessed the relationship between obesity and LVM using the estimated coefficients for BMI in sex-specific linear regressions adjusted
for age and systolic blood pressure, where the outcome was LVM normalized by height. This analysis indicated that obesity
is a significant predictor of LVM conditional on age and systolic
blood pressure for both men and women. 

In order to test whether the assumptions of independence inherently
assumed by \cite{lauer1991impact} are valid, we applied Moran's $I$ to normalized LVM and to BMI, separately for males and females, and to the residuals from our replication of the Lauer et al. sex-specific regressions.  
The results are reported in Table~\ref{tab:LVM_moran}.
If the inference reported in \cite{lauer1991impact} is valid, the errors from the regressions should be independent, however Moran's $I$ provides evidence of network
dependence for the residuals in addition to the marginal LVM variable, for both males and females, undermining the i.i.d. assumption on which the validity of the linear regression model rests. Furthermore, for both sexes there is evidence of network dependence in both LVM and BMI, suggesting that any association may be spurious due to network dependence.

\begin{table}[!ht]
\centering
\caption{\label{tab:LVM_moran} Results of tests of network dependence for males and females, for LVM, BMI, and the residuals from regressing LVM onto covariates. $P$-values are obtained from permutation tests.}
\resizebox{0.8\textwidth}{!}{\begin{tabular}{|l|r|r|}
\hline
$Y$ & $I_{std}$  & $p$-value \\
\thickhline
\textbf{Male} & & \\ \hline
Normalized LVM & 2.26 & 0.01 \\ \hline
BMI &1.64 & 0.05 \\ \hline
Residual from LVM \~{} BMI + age + systolic BP & 1.12 & 0.11 \\
\hline
\textbf{Female} & &\\ \hline
Normalized LVM & 2.23 & 0.02 \\ \hline
BMI & 1.19 & 0.13 \\ \hline
Residual from LVM \~{} BMI + age + systolic BP & 2.08 & 0.03 \\
\hline
\end{tabular}}
\end{table}


Cox proportional hazards models~\citep{cox1992regression}
are commonly applied to the FHS data to assess risk factors for mortality.  When the assumptions of the Cox model hold, including i.i.d. observations, Martingale residuals are expected to be approximately uncorrelated
in finite samples \citep{lin1993checking,tableman2003survival}. We looked for evidence of residual dependence in a study by Tsuji et al. \citep{tsuji1994reduced}  of the association between
eight different heart rate variability (HRV) measures and four-year mortality.
We replicated the twenty-four separate Cox models reported in \cite{tsuji1994reduced}: for each of eight measures of HRV we fit models without adjusting for covariates,
adjusting for age and sex, and adjusting for clinical risk factors in addition to age and sex.

We tested for dependence in two versions of the outcome: survival time ($I_{std} = 3.54$, $p<0.05$) and a binary indicator of death ($I_{std}=1.71$, $p=0.08$), and the top two rows of Table~\ref{tab:moran_table} show the results of tests applied to each exposure of interest.  The strong evidence of dependence in the outcome (censoring notwithstanding) and in some of the exposures suggests that the point estimates from the Tsuji et al. analysis may be subject to spurious associations due to network dependence.  The remainder of Table ~\ref{tab:moran_table} includes the results of tests of dependence applied to the Martingale residuals from the twenty-four
different regression models, which suggest that the i.i.d. assumption
may be violated in most or all of these regressions. Interestingly, Moran's $I$ statistic is larger with smaller $p$-values for the covariates that were found to be significant predictors of all cause mortality.  This is consistent with a hypothesis that the statistically significant associations are spurious, due to network dependence, rather than to true population-level associations.  

\begin{table}[!ht]
	\centering
	\caption{\label{tab:moran_table} Tests of network dependence using Moran's
		$I$ statistic applied to each HRV measure and to the Martingale residuals from the Cox models for eight different HRV measures. $P$-values are obtained from permutation tests.}
	
	\begin{tabular}{|r|c|c|c|c|c|c|c|c|}
		\hline
		HRV measures: & $\ln$SDNN  & $\ln$pNN50  & $\ln$r-MSSD & $\ln$VLF & $\ln$LF  &   $\ln$HF &  $\ln$TP & $\ln$LF/HF \\
	\thickhline
		\multicolumn{9}{|l|}{\textbf{Exposure}} \\  \hline
		$I_{std}$ & 0.33 & -0.41 & -0.12 & 1.72 & 1.62 & 0.83 & 1.85 & -0.03 \\ \hline
		$p$-value & 0.38 & 0.59 & 0.52 & 0.06 & 0.08 & 0.20 & 0.06 & 0.47 \\ \hline
		\multicolumn{9}{|l|}{\textbf{Residuals from unadjusted model for all-cause mortality}} \\ \hline
		$I_{std}$ & 1.57 & 1.65 & 1.64 & 1.38 & 1.38 & 1.54 & 1.38 & 1.59 \\ \hline
		$p$-value & 0.06 & 0.04 & 0.04 & 0.08 & 0.09 & 0.06 & 0.08 & 0.05 \\ \hline
		\multicolumn{9}{|l|}{\textbf{Residuals from model adjusted for age and sex}} \\  \hline
		$I_{std}$ & 1.94 & 2.00 & 2.05 & 1.92 & 1.75 & 1.95 & 1.87 & 1.97 \\ \hline
		$p$-value & 0.02 & 0.02 & 0.02 & 0.02 & 0.04 & 0.02 & 0.03 & 0.03 \\ \hline
		\multicolumn{9}{|l|}{ \textbf{Residuals from model adjusted for age, sex, and clinical risk factors}} \\  \hline
		$I_{std}$ & 1.55 & 1.52 & 1.56 & 1.60 & 1.46 & 1.53 & 1.52 & 1.52 \\  \hline
		$p$-value & 0.07 & 0.07 & 0.07 & 0.06 & 0.09 & 0.07 & 0.09 & 0.07 \\ \hline
	\end{tabular}
\end{table}

\subsection{Peer effects} \label{subsec:peer}

To assess peer influence for obesity using FHS data, ~\cite{christakis2007spread} fit longitudinal logistic regression models of each individual's obesity status at exam $k=2,3,4,5,6,7$  onto each of the individual's social contacts' obesity statuses at exam $k$ and $k-1$ (with a separate entry into the model for each contact), controlling for individual covariates and for the node's own obesity status at exam $k-1$.  They used generalized estimating equations \citep{liang1986longitudinal} to account for correlation within individual, but their model assumes independence across individuals. Christakis and Fowler fit this model separately for ten different types of social connections, including siblings, spouses, and immediate neighbors. 

We replicated a secondary analysis in which the social contacts' obesity statuses at exams $k-1$ and $k-2$ were used instead of $k$ and $k-1$; we replicated this analysis to avoid the misspecification inherent in the former specification \citep{lyons2011spread}.  Although it would be possible to adapt our proposed test of dependence to longitudinal or clustered data, that is beyond the scope of this paper and for simplicity we fit the Christakis and Fowler model at a single time point and selected one social contact for each node in order to have one residual per individual.  We chose to use exam 3 for the outcome data because it gave us the largest sample size.  We looked at sibling relationships because this gives the largest number of ties in the underlying network compared to the other nine types of relationships considered by Christakis and Fowler, and because we had a prior hypothesis that subjects with close genetic relationships would evince dependence in obesity status. 

We calculated Moran's $I$ for the outcome (obesity status in exam 3), the predictor of interest (sibling's obesity status in exam 2), and the residuals from the logistic regression of each node's exam 3 obesity status onto the node's own obesity status in exam 2, the sibling's obesity status in exam 2, the sibling's obesity status at exam 1, and covariates age, sex, and education. For the outcome $I_{std}=7.10$ ($p<0.01$) and for the exposure $I_{std}=15.91$ ($p<0.01$) (because BMI is a binary variable $I$ is equivalent to $\Phi$), suggesting that spurious associations due to network dependence could contribute to any apparent association between the outcome and the exposure of interest. $I_{std}=2.76$ ($p<0.01$) for the regression residuals, providing strong evidence that the i.i.d. assumption on which these analyses rests may be invalid.  Details of our analysis can be found in the Appendix, along a similar analysis for a follow-up paper that used Mendelian randomization to assess peer influence for obesity \citep{o2014estimating}. 

\section{Possible solutions and their limitations}
\label{sec:solutions}

So far we have discussed testing for network dependence, which can help diagnose replication failures or caution against the use if i.i.d. methods, but not correctives for when evidence of dependence is found.  In general, despite increasing interest in and availability of social network data, there is a dearth of valid statistical methods to account for network dependence. There are unsolved challenges for dealing with network dependence, both in terms of valid uncertainty quantification and in terms of correcting spurious correlations due to network dependence. We will explain the challenges and briefly discuss paths forward.

\subsection{Valid standard error estimation}

Although many statistical methods exist for dealing with dependent data, almost all of these methods are intended for spatial or temporal data, or, more broadly, for observations with positions in $\mathbb{R}^{k}$ and dependence that is related to Euclidean distance between pairs of points. The topology of a network is very different from that of Euclidean space, and many of the methods that have been developed to accommodate Euclidean dependence are not appropriate for network dependence. The most important difference is the distribution of pairwise distances which, in Euclidean settings, is usually assumed to skew towards larger distances as the sample grows, with the maximum distance tending to infinity with $n$. In social networks, on the other hand, pairwise distances tend to be concentrated on shorter distances and may be bounded from above.  

Indeed, the first central limit theorems for network dependence were only proven in the last two years, with the working papers that proved them still unpublished at the time of writing \citep{ogburn2017causal,kojevnikov2019limit}.  Both of these central limit theorems place strong restrictions on the nature of network dependence, and it is still an open question when real social network data would meet the required assumptions.  Characterizing when, in real data, we can expect a central limit theorem to approximately hold, and can therefore use approximate normality as a basis of inference, is an active area of research for our group.  Even if we knew that inference under approximate normality were licensed, finding a good standard error estimator, e.g. one that is consistent with the same rate of convergence as the parameter of interest, remains an open challenge in most settings.  If researchers have full knowledge of network structure and of the geodesic distance at which dependence becomes negligible, then it is relatively straightforward to estimate the variance of a sample mean or an M-estimator, because it is clear which of the covariance terms in $var(\bar{Y})=\sum_{i}var(Y_i)+\sum_{i\neq j}cov(Y_i,Y_j)$ can be \emph{a priori} set to $0$.  This is the strategy taken in \cite{ogburn2017causal}.  But in many settings, like the FHS data, we have access to neither the true network structure nor to reasonable information about the distance at which dependence becomes negligible.

If subjects are sampled from many independent social networks, methods for clustered data may be (and often are) used.  Even if only a small number of independent networks are observed, tests may be constructed to compare the difference in estimated means or coefficients between pairs of networks to the expected spread given the estimated standard error, e.g. by subsampling from each cluster.  If a single network is observed, a similar test could be conducted using artificial clusters from different regions in the network.

\subsection{Correcting spurious associations due to network dependence}

Time series and GWAS researchers have been very successful in overcoming the problem of spurious associations due to dependence, but unfortunately network data again proves more challenging.  The strategies that have been successfully employed for times series and GWAS all involve modeling the full dependence structure in order to extract conditionally independent observations, which can then be used to estimate associations without risk of spurious associations due to dependence.  While this same strategy can, in theory, solve the problem of spurious associations in networks, accessing the background knowledge necessary to model the full dependence structure in most network settings is beyond our current ability.  First we will show and discuss some successes in the social network context, then we will highlight the gaps that remain.

Although it departs somewhat from the principle of extracting independent observations conditional on a model of the full dependence structure, there is a growing literature on causal inference in social networks that boasts many successes.  If researchers have conducted a randomized experiment, then hypothesis tests based on the randomization distribution are valid regardless of the dependence structure in the outcomes \citep{rosenbaum2007interference,proschan2008cluster}, and a growing body of literature addresses the design and analysis of network experiments (see \citealp{ogburn2018challenges} for a review and references). Two recent papers have proposed methods for dealing with network structure in observational causal inference \citep{ogburn2017causal,tchetgen2017auto}.  However, both require observing all network ties in addition to strong assumptions limiting the nature of dependence.   

When some network ties are familial, and when genetic data is available, as is the case in the FHS, techniques developed to control for confounding due to cryptic relatedness~\citep{sillanpaa2011overview} may be helpful for estimating the unknown familial network structure and for controlling for confounding due to that structure.  Similarly, if researchers have \emph{a priori} knowledge of a model for the dependence structure, then it could be straightforward to correct standard error estimators for dependence and to control for dependence before estimating associations.   For example, if observations are independent conditional on observed features of the network or covariates associated with neighboring nodes, then any analysis that conditions on those random variables will be valid; and maximum likelihood methods may be appropriate if the data are distributed according to a known parametric distribution with a sparse covariance matrix of known parametric form.  

\begin{figure}[ht]
	\centering
	\includegraphics[width=\linewidth]{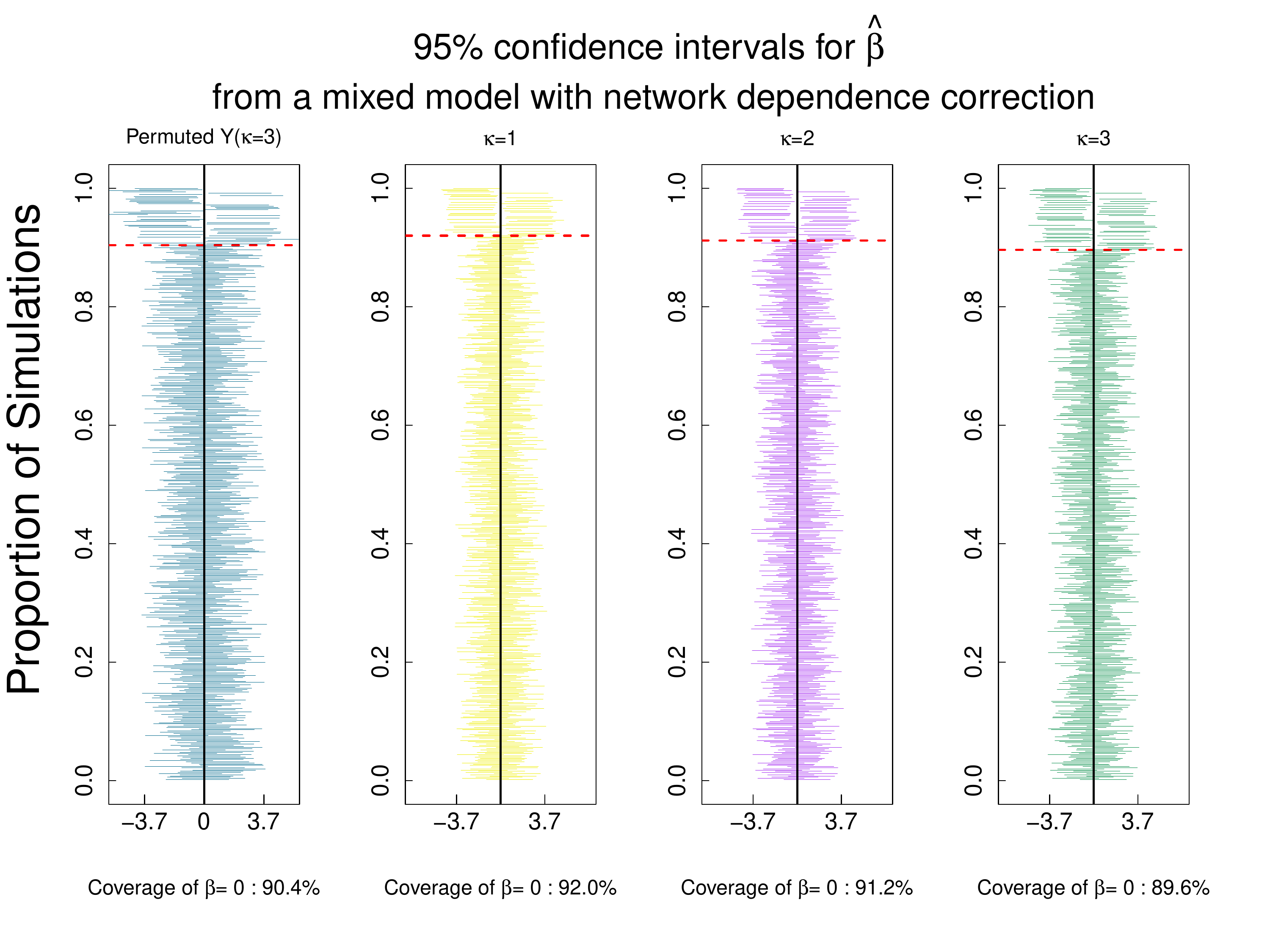}
	\caption{Each column contains 95$\%$ confidence intervals (CIs) for the $X$ coefficient, $\beta$, from a mixed model regressing $Y$ on 	
		$X$, using the true covariance matrix as a "kinship" matrix. 
		$X$ and $Y$ were simulated with increasing dependence from left (no dependence) to right. The CIs above the dotted
		line do not contain the true $\beta=0$ (red-line) while the CIs below
		the dotted line contain $0$. Coverage rates of 95$\%$ CIs are
		calculated as the percentages of the CIs covering $0$.}
	\label{fig:spurious_correction}
\end{figure}

As an example, we reanalyzed the data in Figure \ref{fig:confounding} using a mixed model that was proposed by \cite{sul2018population} to correct for population structure and cryptic relatedness in GWAS studies.  This method, like others that are commonly used in GWAS studies, requires as an input the \emph{kinship matrix}, which determines the relatedness of the individuals in the study and therefore fully describes the dependence structure.  When we implemented the \cite{sul2018population} method using, in place of a kinship matrix, the true covariance matrix describing the dependence across network nodes, we were able to largely correct for spurious associations due to network dependence.  Figure \ref{fig:spurious_correction} shows our reanalysis of the simulation data in Figure \ref{fig:confounding} where, instead of assuming the data are i.i.d., we used a mixed model with the correct covariance matrix.  However we found, as is expected, that this method is very sensitive to the correct specification of the covariance (kinship) matrix; if this matrix is even slightly misspecified, coverage rates drop. This can be seen in Figure A3 in the Appendix.  

In general, we are not aware of any methods that can avoid spurious associations due to dependence without complete understanding of the dependence structure. In our simulated data we knew the covariance matrix exactly, but in real social networks, especially if dependence may be due to latent variables and to direct transmission, it is difficult to imagine being able to \emph{a priori} specify a correct model for the dependence.  For example, in the FHS data the observed adjacency matrix may be the best information we have about the dependence structure in the data.  We reanalyzed \cite{lauer1991impact} using the mixed model approach above, substituting the adjacency matrix for the kinship matrix, and tested for network dependence after the correction.  We found that a few of the covariates had attenuated test statistics, indicating that controlling for the dependence captured by adjacency matrix may have had a benefit, but the tests still showed strong evidence of network dependence, suggesting that this strategy largely failed to account for the dependence in the data.  Table A9 in the Appendix shows the side-by-side comparison of the analysis with and without a correction based on the adjacency matrix.

Future work is needed to flesh out the proposals above, to develop methods to account for network dependence when the network is partially observed, and to develop methods for settings more general than the ones mentioned above, all of which involve structure or assumptions beyond what we would expect in the typical network dependence setting.  Because it is difficult to imagine being able to \emph{a priori} specify a correct model for the dependence in most social network settings, methods to learn about the dependence from the data are especially urgent.

\section{Discussion}
\label{sec:conclusion}
As researchers across many scientific disciplines grapple with replication crises, many sources of artificially small $p$-values and increased false positive rates have received attention, but the possible impact of network dependence has been overlooked. In this paper, we used simple tests for independence among observations sampled from a single network to demonstrate that many types of analyses using FHS data may have reported error-prone point estimates and artificially small $p$-values, standard errors, and confidence intervals due to unacknowledged network dependence.

Tests for network dependence rely on social network information, which, as we have noted, is not available in most studies that are not explicitly about networks. However, missing data on network ties will generally affect the power but not validity of these tests, so adding information on even just one or two ties per subject to a data collection protocol would enable researchers to test for network dependence. 

Beyond a call for methods development, our primary recommendation to researchers designing new studies with human subjects is to avoid recruiting from one or a small number of underlying social networks whenever possible, especially if an outcome or exposure of interest could plausibly exhibit network dependence. 
Researchers working with existing data should be aware of the possibility that social network dependence may undermine the use of i.i.d. models.

\section*{Acknowledgements}

 The authors were supported by ONR grants N000141512343 and N000141812760. The Framingham Heart Study is conducted and supported by the National Heart, Lung, and Blood Institute (NHLBI) in collaboration with Boston University (Contract No. N01-HC-25195 and HHSN268201500001I). This manuscript was not prepared in collaboration with investigators of the Framingham Heart Study and does not necessarily reflect the opinions or views of the Framingham Heart Study, Boston University, or NHLBI. The authors are grateful to Caroline Epstein, whose M.S. thesis this work builds upon, and to Rob Kass, Don Geman, Andrew Gelman, Marshall Joffe, Tom Louis, James O'Malley, Jamie Robins, Cosma Shalizi, Eric Tchetgen Tchetgen, Alex Volfovsky, Nathan Winkler-Rhoades, two anonymous reviewers, members of the University of Pennsylvania Causal Inference Reading Group, and members of the University of Chicago Econometrics and Statistics Group for helpful comments.

\bibliographystyle{jasa}
\bibliography{reference}

\end{document}